Letter

# Measurement of the $B^+/B^0$ production ratio in $e^+e^-$ collisions at the $\Upsilon(4S)$ resonance using $B \to J/\psi(\ell\ell)K$ decays at Belle

S. Choudhury, S. Sandilya, K. Trabelsi, A. Giri, I. Adachi, H. Aihara, D. M. Asner, H. Atmacan, T. Aushev, R. Ayad, V. Babu, S. Bahinipati, P. Behera, K. Belous, J. Bennett, M. Bessner, V. Bhardwaj, B. Bhuyan, T. Bilka, D. Bodrov, G. Bonvicini, J. Borah, A. Bozek, M. Bračko, P. Branchini, T. E. Browder, A. Budano, M. Campajola, D. Červenkov, M.-C. Chang, C. Chen, B. G. Cheon, K. Chilikin, H. E. Cho, K. Cho, S.-J. Cho, S.-K. Choi, Y. Choi, D. Cinabro, S. Das, G. De Pietro, F. Di Capua, J. Dingfelder, Z. Doležal, T. V. Dong, D. Epifanov, D. Ferlewicz, B. G. Fulsom, R. Garg, V. Gaur, N. Gabyshev, P. Goldenzweig, E. Graziani, T. Gu, K. Gudkova, C. Hadjivasiliou, T. Hara, K. Hayasaka, H. Hayashii, M. T. Hedges, D. Herrmann, W.-S. Hou, C.-L. Hsu, K. Inami, N. Ipsita, A. Ishikawa, R. Itoh, M. Iwasaki, W. W. Jacobs, E.-J. Jang, Q. P. Ji, S. Jia, Y. Jin, K. K. Joo, K. H. Kang, T. Kawasaki, C. Kiesling, C. H. Kim, D. Y. Kim, K.-H. Kim, Y.-K. Kim, H. Kindo, K. Kinoshita, P. Kodyš, T. Konno, A. Korobov, S. Korpar, E. Kovalenko, P. Križan, P. Krokovny, M. Kumar, R. Kumar, K. Kumara, Y.-J. Kwon, T. Lam, J. S. Lange, M. Laurenza, S. C. Lee, P. Lewis, J. Li, L. K. Li, Y. B. Li, L. Li Gioi, J. Libby, K. Lieret, D. Liventsev, T. Luo, M. Masuda, T. Matsuda, D. Matvienko, S. K. Maurya, F. Meier, M. Merola, F. Metzner, K. Miyabayashi, R. Mizuk, G. B. Mohanty, I. Nakamura, M. Nakao, Z. Natkaniec, A. Natochii, L. Nayak, M. Nayak, N. K. Nisar, S. Nishida, S. Ogawa, H. Ono, P. Oskin, P. Pakhlov, G. Pakhlova, S. Pardi, S.-H. Park, A. Passeri, S. Patra, S. Paul, T. K. Pedlar, R. Pestotnik, L. E. Piilonen, T. Podobnik, S. Prell, E. Prencipe, M. T. Prim, N. Rout, G. Russo, Y. Sakai, A. Sangal, L. Santelj, V. Savinov, G. Schnell, J. Schueler, C. Schwanda, A. J. Schwartz, Y. Seino, K. Senyo, M. E. Sevior, M. Shapkin, C. Sharma, C. P. Shen, J.-G. Shiu, J. B. Singh, A. Sokolov, E. Solovieva, M. Starič, Z. S. Stottler, J. F. Strube, M. Sumihama, T. Sumiyoshi, M. Takizawa, K. Tanida, F. Tenchini, M. Uchida, T. Uglov, Y. Unno, S. Uno, P. Urquijo, Y. Ushiroda, Y. Usov, R. van Tonder, G. Varner, K. E. Varvell, A. Vinokurova, A. Vossen, E. Waheed, E. Wang, M.-Z. Wang, X. L. Wang, M. Watanabe, S. Watanuki, E. Won, X. Xu, B. D. Yabsley, S. B. Yang, J. Yelton, J. H. Yin, C. Z. Yuan, Y. Yusa, Y. Zhai, Z. P. Zhang, V. Zhilich, and V. Zhukova

(Belle Collaboration)

We measure the ratio of branching fractions for the $\Upsilon(4S)$ decays to $B^+B^-$ and $B^0\bar{B}^0$ using $B^+ \to J/\psi(\ell\ell)K^+$ and $B^0 \to J/\psi(\ell\ell)K^0$ samples, where $J/\psi(\ell\ell)$ stands for $J/\psi \to \ell^+\ell^-$ ($\ell = e$ or $\mu$), with 711 $\mathrm{fb}^{-1}$ of data collected at the $\Upsilon(4S)$ resonance with the Belle detector. We find the decay rate ratio of $\Upsilon(4S) \to B^+B^-$ over $\Upsilon(4S) \to B^0\bar{B}^0$ to be $1.065 \pm 0.012 \pm 0.019 \pm 0.047$, which is the most precise measurement to date. The first and second uncertainties are statistical and systematic, respectively, and the third uncertainty is due to the assumption of isospin symmetry in $B \to J/\psi(\ell\ell)K$.

The production ratio $R^{+/0}$ of $B^+/B^0$ at the $\Upsilon(4S)$ resonance is given by the ratio of decay rates of $\Upsilon(4S) \to B^+B^-$ to $\Upsilon(4S) \to B^0\bar{B}^0$,

$$R^{+/0} = f^{+-}/f^{00} = \frac{\Gamma[\Upsilon(4S) \to B^+B^-]}{\Gamma[\Upsilon(4S) \to B^0\bar{B}^0]}. \quad (1)$$

A precise knowledge of this observable is essential for the measurement of decay branching fractions of charged and neutral $B$ mesons produced in $\Upsilon(4S)$ decays. It is also important for extracting the isospin asymmetry, $A_I = \frac{(\tau_{B^+}/\tau_{B^0})\mathcal{B}[B^0\to K^0\ell^+\ell^-]-\mathcal{B}[B^+\to K^+\ell^+\ell^-]}{(\tau_{B^+}/\tau_{B^0})\mathcal{B}[B^0\to K^0\ell^+\ell^-]+\mathcal{B}[B^+\to K^+\ell^+\ell^-]}$ ($\ell = e$ or $\mu$), where $\tau_{B^+}/\tau_{B^0}$ denotes the ratio of $B^+$ and $B^0$ lifetimes, to search for new physics [1–3]. As the charged and neutral $B$ mesons have very similar masses [$5279.34 \pm 0.12$ *vs.* $5279.65 \pm 0.12\ \mathrm{MeV}/c^2$], one would naively expect the ratio to be close to unity. However, Coulomb corrections could be as large as 20% [4]. Other calculations, based on nonrelativistic effective field theory or taking into account the structure of mesons as well as final-state interactions, predict values for the ratio in the ranges 1.1–1.2 [5] and 0.9–1.2 [6], respectively.

Two complementary approaches have been used to measure $R^{+/0}$ or $f^{+-}/f^{00}$, both of them so far providing results that are consistent with unity:

(i) With the assumption of isospin symmetry:

Two decay channels have been used, for which the isospin symmetry is expected to hold well:

(a) Measurement of the partial decay width for the semileptonic decays $\bar{B} \to D^* \ell^- \bar{\nu}_\ell$. The $R^{+/0}$ value measured by the CLEO Collaboration [7] is $1.058 \pm 0.084 \pm 0.136$. Hereafter, the first and second quoted uncertainties are statistical and systematic, respectively. Using 29 $\mathrm{fb}^{-1}$ of data, Belle [8] found $R^{+/0} = 1.01 \pm 0.03 \pm 0.09$.

(b) Using $B \to (c\bar{c})K^{(*)}$ [9–11] charmonium decays. The last and most precise measurement of this observable was from *BABAR* [9], $R^{+/0} = 1.06 \pm 0.02 \pm 0.03$, with 112.4 $\mathrm{fb}^{-1}$ of data. Here, $R^{+/0}$ is calculated from the measured branching fractions of $B^+$ and $B^0$ decays.

(ii) Without the assumption of isospin symmetry:

The value of $f^{00}$ has been measured by *BABAR* [12] in $\bar{B}^0 \to D^{*+} \ell^- \bar{\nu}_\ell$ decays using single and double reconstruction without the isospin assumption. A similar study using the full Belle data sample would provide a statistically less precise $R^{+/0}$ result compared to the one based on $B \to J/\psi(\ell\ell)K$ decays presented in this paper. Here, $J/\psi(\ell\ell)$ stands for $J/\psi \to \ell^+\ell^-$.

Results of the branching fractions and isospin asymmetry of the $B^{\pm,0} \to J/\psi(\ell\ell)K^{\pm,0}$ decay channels have been reported by the Belle Collaboration [13]. The measurements were done on a data sample of 711 $\mathrm{fb}^{-1}$ recorded at the KEKB $e^+e^-$ collider [14] operating at the $\Upsilon(4S)$ resonance. In this paper, we present a measurement of $R^{+/0}$ using the same decay channels and data sample. With the event selection specially optimized for $B^{\pm,0} \to J/\psi(\ell\ell)K^{\pm,0}$ decays, this analysis is more robust than the one in Ref. [13]. These decays are good candidates for measuring $R^{+/0}$ since the possible contribution of isospin symmetry breaking from rescattering in $B \to J/\psi(\ell\ell)K$ is expected to be small in the Standard Model, of the order of $\bar{\lambda}^3$ [15]. Here $\bar{\lambda} \simeq 0.2$ [16,17], which is of the same order as the Wolfenstein parameter $\lambda$ [18]. In the presence of very large rescattering effects, symmetry breaking corrections could be as large as $\mathcal{O}(\bar{\lambda}^2)$, but there is currently no evidence for this [15]. QCD factorization approaches also do not support such large effects [19,20].

The yields of $B^+ \to J/\psi(\ell\ell)K^+$ and $B^0 \to J/\psi(\ell\ell)K^0$ [21] are given as

$$N_{\mathrm{sig}}^{+} = 2N_{B\bar{B}} f^{+-} \varepsilon^{+} \mathcal{B}[B^+ \to J/\psi(\ell\ell)K^+], \tag{2}$$

$$N_{\mathrm{sig}}^{0} = 2N_{B\bar{B}} f^{00} \varepsilon^{0} \mathcal{B}[B^0 \to J/\psi(\ell\ell)K^0], \tag{3}$$

where $N_{\mathrm{sig}}^{+}$, $N_{\mathrm{sig}}^{0}$, $\varepsilon^+$, and $\varepsilon^0$ are the signal yields and reconstruction efficiencies of charged and neutral $B$ mesons, respectively; $N_{B\bar{B}}$ is the number of $B\bar{B}$ pairs, i.e., $772 \times 10^6$. This leads to

$$\begin{aligned} \frac{N_{\mathrm{sig}}^{+}/\varepsilon^{+}}{N_{\mathrm{sig}}^{0}/\varepsilon^{0}} &= R^{+/0} \frac{\mathcal{B}[B^+ \to J/\psi(\ell\ell)K^+]}{\mathcal{B}[B^0 \to J/\psi(\ell\ell)K^0]} \\ &= R^{+/0} \frac{\Gamma[B^+ \to J/\psi(\ell\ell)K^+]\tau^+}{\Gamma[B^0 \to J/\psi(\ell\ell)K^0]\tau^0}, \\ \Rightarrow R^{+/0} &= \frac{N_{\mathrm{sig}}^{+}\varepsilon^0 \tau_0}{N_{\mathrm{sig}}^{0}\varepsilon^+ \tau_+}, \end{aligned} \tag{4}$$

assuming isospin invariance in $B \to J/\psi(\ell\ell)K$, i.e., $\Gamma[B^+ \to J/\psi(\ell\ell)K^+] = \Gamma[B^0 \to J/\psi(\ell\ell)K^0]$ or $\mathcal{B}[B^+ \to J/\psi(\ell\ell)K^+]\tau_0 = \mathcal{B}[B^0 \to J/\psi(\ell\ell)K^0]\tau_+$. The value used for the ratio $\tau_+/\tau_0$ is $1.076 \pm 0.004$ [22].

The Belle detector [23] is a large-solid-angle magnetic spectrometer composed of a silicon vertex detector (SVD), a 50-layer central drift chamber (CDC), an array of aerogel threshold Cherenkov counters (ACC), a barrel-like arrangement of time-of-flight scintillation counters (TOF), and an electromagnetic calorimeter (ECL) comprising CsI(Tl) crystals. All of these subdetectors are located inside a superconducting solenoid coil that provides a 1.5 T magnetic field. An iron flux-return yoke placed outside the coil is instrumented with resistive plate chambers (KLM) to detect $K_L^0$ mesons and muons. Two inner detector configurations are used: a 2.0 cm radius beam pipe and a three-layer SVD for the first sample of 140 $\mathrm{fb}^{-1}$; and a 1.5 cm radius beam pipe, a four-layer SVD, and a small-inner-cell CDC for the remaining 571 $\mathrm{fb}^{-1}$ [24].

We use Monte Carlo (MC) simulated events to study the properties of signal decays and to identify various background sources. The $B \to J/\psi(\ell\ell)K$ decays are generated with the EvtGen package [25] using the scalar to vector and scalar (SVS) model for $B \to J/\psi K$ decays, and the vector to lepton and lepton (VLL) model for $J/\psi \to \ell^+\ell^-$ decays. The PHOTOS package [26] is used to incorporate final-state radiation effects, while GEANT3 [27] is used for detector simulation.

We reconstruct $B^+ \to J/\psi(\ell\ell)K^+$ and $B^0 \to J/\psi(\ell\ell)K_S^0$ decays. The charged particles originating from the interaction point (IP) are selected, except for daughters of $K_S^0$, by requiring their impact parameters to be less than 4.0 cm along the $z$ axis (direction opposite to the $e^+$ beam) and less than 1.0 cm in the transverse plane. We apply a minimum momentum threshold of 100 MeV to reduce the background from low-momentum particles. The kaon candidates are selected using a likelihood ratio, $\mathcal{R}_{K/\pi} = \mathcal{L}_K/(\mathcal{L}_K + \mathcal{L}_\pi)$, where $\mathcal{L}_K$ and $\mathcal{L}_\pi$ are the likelihoods of the charged particle being a kaon or a pion, respectively. The likelihoods are calculated based on the number of photoelectrons in the ACC, the specific ionization in the CDC, and the flight time in the TOF. We select the charged particles that satisfy $\mathcal{R}_{K/\pi} > 0.6$, which results in a 90% kaon selection efficiency with a 9% pion misidentification rate. The $K_S^0 \to \pi^+\pi^-$ decays are identified with a neural network [28] comprising the following 13 variables: the $K_S^0$ momentum in the lab frame; the separation along the $z$ axis between the two $\pi^\pm$ tracks; the impact parameter with respect to the IP transverse to the $z$ axis of each $\pi^\pm$ track; the $K_S^0$ flight length in the transverse plane; the angle between the $K_S^0$ momentum and the vector joining the IP and the $K_S^0$ vertex; the angle between the $\pi^+$ momentum and the lab frame $e^+e^-$ boost direction, in the $K_S^0$ rest frame; the number of CDC hits in both stereo and axial views for each $\pi^\pm$ track; and the presence or absence of SVD hits for each $\pi^\pm$ track. We also require that the reconstructed $K_S^0$ invariant mass be between 487 and 508 $\mathrm{MeV}/c^2$, which is $\pm 3\sigma$ around the nominal mass [22]. The $K_S^0$ reconstruction efficiency is approximately 82% [29]. The muon candidates are selected based on information from the KLM, requiring a muon likelihood ratio $\mathcal{R}_\mu = \mathcal{L}_\mu/(\mathcal{L}_\mu + \mathcal{L}_K + \mathcal{L}_\pi) > 0.9$, where $\mathcal{L}_\mu$ is the muon likelihood value. This criterion results in an efficiency of 89% with a pion misidentification rate of 1.5% [30]. The minimum momentum is required to be 0.8 $\mathrm{GeV}/c$ in order to ensure the muon candidates reach the KLM. The electron candidates are required to have a minimum momentum of 0.5 $\mathrm{GeV}/c$ and electron likelihood ratio $\mathcal{R}_e = \mathcal{L}_e/(\mathcal{L}_e + \mathcal{L}_{\tilde{e}}) > 0.9$, where $\mathcal{L}_e$ and $\mathcal{L}_{\tilde{e}}$ are the likelihood values for electron and nonelectron hypotheses, respectively. These likelihoods are calculated using the ratio of calorimetric cluster energy to the track momentum; the shower shape in the ECL; the matching of the track with the ECL cluster; the specific ionization in the CDC; and the number of photoelectrons in the ACC [31]. The electron selection efficiency is 92% with a less than 1% pion misidentification rate. The energy loss due to bremsstrahlung is recovered by considering the photons found in a 50 mrad cone along the initial momentum direction of the electron.

Two oppositely charged leptons are combined with a $K^+$ or $K_S^0$ candidate to form a $B^+$ or $B^0$ meson. The invariant mass requirements for the $J/\psi \to \mu\mu$ and $J/\psi \to ee$ channels are $2.95 < M_{\mu\mu} < 3.18\ \mathrm{GeV}/c^2$ and $2.85 < M_{ee} < 3.18\ \mathrm{GeV}/c^2$, respectively. The kinematic variables that distinguish signal from background are the beam-energy constrained mass $M_{\mathrm{bc}}$ and the energy difference $\Delta E$, and are given by

$$M_{\mathrm{bc}} = \sqrt{(E_{\mathrm{beam}}/c^2)^2 - (p_B/c)^2},$$
$$\Delta E = E_B - E_{\mathrm{beam}},$$

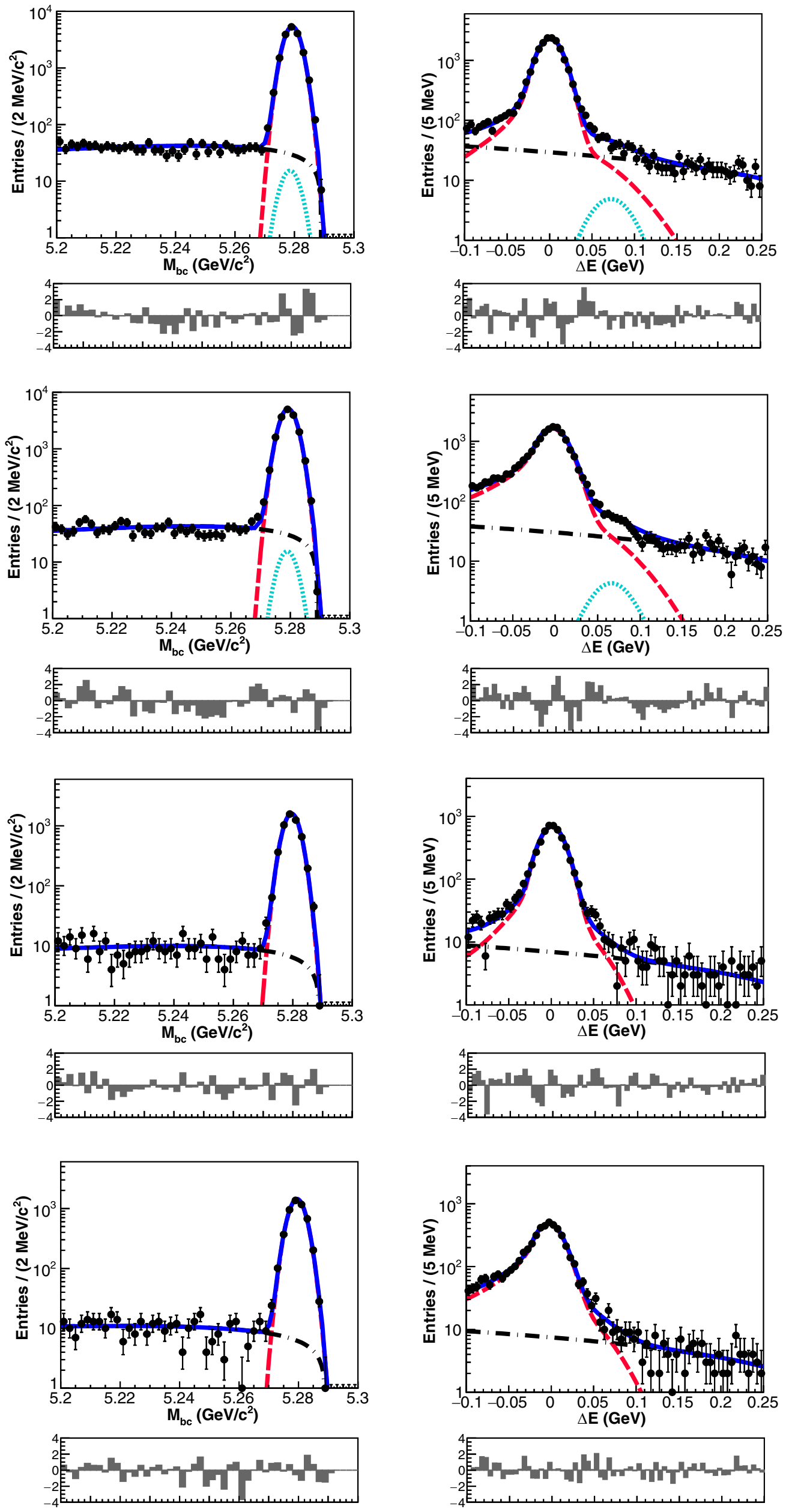

FIG. 1. Left (right) panels: $M_{\mathrm{bc}}$ ($\Delta E$) projections of two-dimensional unbinned extended maximum-likelihood fits to the data events that pass the selection criteria for $B^+ \to J/\psi(\mu\mu)K^+$ (first row), $B^+ \to J/\psi(ee)K^+$ (second row), $B^0 \to J/\psi(\mu\mu)K_S^0$ (third row), and $B^0 \to J/\psi(ee)K_S^0$ (fourth row) in the log scale. Points with error bars are the data; blue solid curves are the fitted results for the signal-plus-background hypothesis; red dashed curves denote the signal component; black dash-dotted and cyan dotted are combinatorial and $[\pi^+ J/\psi]$ backgrounds, respectively. The lower plots show the deviation of data from the fit prediction normalized by the data uncertainty.

where $E_{\mathrm{beam}}$ is the beam energy, and $E_B$ and $p_B$ are, respectively, the energy and momentum of the $B$ candidate. These quantities are calculated in the $e^+e^-$ center-of-mass (c.m.) frame. The candidates that satisfy $M_{\rm bc} > 5.20\ {\rm GeV}/c^2$ and $-0.10 < \Delta E < 0.25$ GeV are retained for further analysis.

With the above selection criteria applied, less than 2% of signal MC events are found to have more than one $B$ candidate. For these events, we retain the candidate with the smallest $\chi^2$ value obtained from a vertex fit of the $B$ decay products. From MC simulation, this criterion is found to select the correct signal candidate 78%–85% of the time, depending on the decay mode. The candidates arising from $B^0 \to J/\psi K^{*0}$ populate the negative side of $\Delta E$ and are removed with the $\Delta E > -0.10$ GeV criterion.

The signal yield is extracted by performing a two-dimensional unbinned extended maximum-likelihood fit to the $M_{\rm bc}$ and $\Delta E$ distributions. The signal component of $M_{\rm bc}$ is modeled with a Gaussian function. Its $\Delta E$ shape is modeled with the sum of a Crystal Ball [32] and a Gaussian function with a common mean. The $M_{\rm bc}$ and $\Delta E$ distributions of the combinatorial background are modeled with an ARGUS function [33] and a first-order Chebyshev polynomial, respectively. We find that the $B^+ \to J/\psi(\ell\ell)\pi^+$ background contributes to the $B^+ \to J/\psi(\ell\ell)K^+$ channel. The shape parameters of this background along with its yield divided by that of $B^+ \to J/\psi(\ell\ell)K^+$, using the pion-to-kaon fake rate from data, are fixed in the fit. The signal parameters of the fit in data are fixed to those in signal MC events with free width scale factors and mean shifts. All background parameters are floated in the fit. The fits are performed separately for charged and neutral $B$ decay channels containing $\mu\mu$ and $ee$ in the final states. The fit results are shown in Fig. 1.

There are $17536 \pm 133$, $17193 \pm 132$, $5150 \pm 72$, and $4805 \pm 70$ signal events for $B^+ \to J/\psi(\mu\mu)K^+$, $B^+ \to J/\psi(ee)K^+$, $B^0 \to J/\psi(\mu\mu)K^0_S$, and $B^0 \to J/\psi(ee)K^0_S$ channels, respectively, having a signal purity of greater than 90%. Using Eqs. (2) and (3) along with the signal yields and efficiencies listed in Table I, we get

$$2f^{+-}\mathcal{B}[B^+ \to J/\psi K^+] = (1.024 \pm 0.007 \pm 0.020) \times 10^{-3},$$
$$2f^{00}\mathcal{B}[B^0 \to J/\psi K^0] = (0.892 \pm 0.010 \pm 0.023) \times 10^{-3},$$

averaging $J/\psi \to \mu^+\mu^-$ and $J/\psi \to e^+e^-$ channels. Here, we take $\mathcal{B}[B^0 \to J/\psi(\ell\ell)K^0] = 2 \times \mathcal{B}[B^0 \to J/\psi(\ell\ell)K^0_S]$ and the value of $\mathcal{B}[J/\psi \to \ell\ell]$ from Ref. [22]. Given independent estimates of $f^{+-}$ and $f^{00}$, these results can be used to extract the branching fractions for $B^{\pm,0} \to J/\psi K^{\pm,0}$. In this analysis, we take these branching fractions as inputs in order to obtain a measurement of $R^{+/0}$. Inserting the signal yields and efficiencies from Table I in Eq. (4), we obtain

$$R^{+/0}(\mu\mu) = 1.068 \pm 0.017 \pm 0.019 \pm 0.047,$$
$$R^{+/0}(ee) = 1.062 \pm 0.017 \pm 0.019 \pm 0.047,$$
$$R^{+/0}(\text{avg}) = 1.065 \pm 0.012 \pm 0.019 \pm 0.047.$$

The first and second uncertainties are statistical and systematic, respectively, while the third uncertainty is due to the isospin symmetry assumption in $B \to J/\psi(\ell\ell)K$ decays.

There are several sources of systematic uncertainties contributing to the measurement of $R^{+/0}$. The systematic uncertainty due to kaon identification is found to be 0.8% from a study of the $D^{*+} \to D^0(K^-\pi^+)\pi^+$ sample. The $K^0_S$ identification uncertainty is 1.6% [29]. The systematic uncertainty due to charged track reconstruction is 0.35% per track estimated by using the partially reconstructed $D^{*+} \to D^0\pi^+$, $D^0 \to \pi^-\pi^+K^0_S$, and $K^0_S \to \pi^+\pi^-$ events. There are three (four) charged tracks for charged (neutral) $B$ channels, and thus in the ratio of $R^{+/0}$ the track reconstruction uncertainty is 0.35% due to an additional track in the neutral $B$ case. The uncertainty in efficiency because of limited MC statistics is less than 0.2%. The shape parameters fixed in the fit are varied by $\pm1\sigma$ from their mean values, and the deviation from the nominal fit value of $N_{\rm sig}$ is the uncertainty due to the signal and background shapes; this is found to be negligible. We take $\bar{\lambda}^3 = 1.1\%$ as the uncertainty due to the isospin symmetry assumption in the decay amplitude, which leads to a 4.4% uncertainty in $R^{+/0}$.

The additional sources of systematic uncertainty contributing to the $2f^{+-(00)}\mathcal{B}[B^{\pm(0)} \to J/\psi K^{\pm(0)}]$ measurement are lepton identification, number of $B\bar{B}$ pairs, and track reconstruction efficiency. The muon and electron identification uncertainties are 0.3% and 0.4% per lepton, respectively. The uncertainties on the track reconstruction efficiency are 1.05% and 1.40% for charged and neutral $B$ decay modes, respectively. The uncertainty due to the number of $B\bar{B}$ pairs is 1.4% [13]. Furthermore, systematic uncertainties (from $\mathcal{B}[J/\psi \to \ell\ell]$) of 0.55% and 0.54% are attributed for modes involving muons and electrons in the final state, respectively. The individual sources of uncertainties are assumed to be independent and are added in quadrature to arrive at the total uncertainty.

TABLE I. Results for $B^{\pm,0} \to J/\psi(\ell\ell)K^{\pm,0}$. The columns correspond to the decay channel, the signal yield from fit ($N_{\rm sig}$), efficiency corrected for data-MC differences ($\varepsilon$), and $2\,f^{+-(00)}\,\mathcal{B}[B^{\pm,0} \to J/\psi(\ell\ell)K^{\pm,0}]$.

| Decay mode | $N_{\rm sig}$ | $\varepsilon$ (%) | $2f^{+-(00)}\ \mathcal{B}[B \to J/\psi(\ell\ell)K]$ |
|---|---|---|---|
| $B^+ \to J/\psi(\mu\mu)K^+$ | $17536 \pm 133$ | 36.9 | $(6.16 \pm 0.04 \pm 0.12) \times 10^{-5}$ |
| $B^+ \to J/\psi(ee)K^+$ | $17193 \pm 132$ | 36.8 | $(6.06 \pm 0.04 \pm 0.12) \times 10^{-5}$ |
| $B^0 \to J/\psi(\mu\mu)K^0_S$ | $5150 \pm 72$ | 24.9 | $(2.68 \pm 0.04 \pm 0.07) \times 10^{-5}$ |
| $B^0 \to J/\psi(ee)K^0_S$ | $4805 \pm 70$ | 23.5 | $(2.65 \pm 0.04 \pm 0.07) \times 10^{-5}$ |

In summary, we have measured the $B^+/B^0$ production ratio at the $\Upsilon(4S)$ resonance using $B^+ \to J/\psi(\ell\ell)K^+$ and $B^0 \to J/\psi(\ell\ell)K_S^0$ decays with the full Belle data sample of 711 $\mathrm{fb}^{-1}$. The observed value of $1.065 \pm 0.012 \pm 0.019 \pm 0.047$ is the most precise measurement to date and is consistent with the world average [34] of $1.059 \pm 0.027$. This result will help to reduce the overall systematic uncertainty for all charged and neutral $B$ decay branching fraction measurements with the $B$ meson coming from $\Upsilon(4S)$ decays. In addition, the branching fraction products $2f^{+-}\mathcal{B}[B^+ \to J/\psi K^+]$ and $2f^{00}\mathcal{B}[B^0 \to J/\psi K^0]$ are measured to be $(1.024 \pm 0.007 \pm 0.020) \times 10^{-3}$ and $(0.892 \pm 0.010 \pm 0.023) \times 10^{-3}$, respectively, which supersede previous results [13].

## ACKNOWLEDGMENTS

This work, based on data collected using the Belle detector, which was operated until June 2010, was supported by the Ministry of Education, Culture, Sports, Science, and Technology (MEXT) of Japan, the Japan Society for the Promotion of Science (JSPS), and the Tau-Lepton Physics Research Center of Nagoya University; the Australian Research Council including Grants No. DP180102629, No. DP170102389, No. DP170102204, No. DE220100462, No. DP150103061, and No. FT130100303; Austrian Federal Ministry of Education, Science and Research (FWF) and FWF Austrian Science Fund No. P 31361-N36; the National Natural Science Foundation of China under Contracts No. 11675166, No. 11705209, No. 11975076, No. 12135005, No. 12175041, and No. 12161141008; Key Research Program of Frontier Sciences, Chinese Academy of Sciences (CAS), Grant No. QYZDJ-SSW-SLH011; Project No. ZR2022JQ02 supported by Shandong Provincial Natural Science Foundation; the Ministry of Education, Youth and Sports of the Czech Republic under Contract No. LTT17020; the Czech Science Foundation Grant No. 22-18469S; Horizon 2020 ERC Advanced Grant No. 884719 and ERC Starting Grant No. 947006 "InterLeptons" (European Union); the Carl Zeiss Foundation, the Deutsche Forschungsgemeinschaft, the Excellence Cluster Universe, and the VolkswagenStiftung; the Department of Atomic Energy (Project Identification No. RTI 4002) and the Department of Science and Technology of India; the Istituto Nazionale di Fisica Nucleare of Italy; National Research Foundation (NRF) of Korea Grants No. 2016R1D1A1B02012900, No. 2018R1A2B3003643, No. 2018R1A6A1A06024970, No. RS202200197659, No. 2019R1I1A3A01058933, No. 2021R1A6A1A03043957, No. 2021R1F1A1060423, No. 2021R1F1A1064008, and No. 2022R1A2C1003993; Radiation Science Research Institute, Foreign Large-size Research Facility Application Supporting project, the Global Science Experimental Data Hub Center of the Korea Institute of Science and Technology Information and KREONET/GLORIAD; the Polish Ministry of Science and Higher Education and the National Science Center; the Ministry of Science and Higher Education of the Russian Federation, Agreement No. 14.W03.31.0026, and the HSE University Basic Research Program, Moscow; University of Tabuk Research Grants No. S-1440-0321, No. S-0256-1438, and No. S-0280-1439 (Saudi Arabia); the Slovenian Research Agency Grants No. J1-9124 and No. P1-0135; Ikerbasque, Basque Foundation for Science, Spain; the Swiss National Science Foundation; the Ministry of Education and the Ministry of Science and Technology of Taiwan; and the U.S. Department of Energy and the National Science Foundation. These acknowledgements are not to be interpreted as an endorsement of any statement made by any of our institutes, funding agencies, governments, or their representatives. We thank the KEKB group for the excellent operation of the accelerator; the KEK cryogenics group for the efficient operation of the solenoid; the KEK computer group and the Pacific Northwest National Laboratory (PNNL) Environmental Molecular Sciences Laboratory (EMSL) computing group for strong computing support; and the National Institute of Informatics, and Science Information NETwork 6 (SINET6) for valuable network support.